\documentclass[aps,nofootinbib,preprint,showkeys]{revtex4}
\usepackage{amsmath}
\usepackage{amsfonts}
\usepackage{amssymb}
\usepackage{bm}
\usepackage{dcolumn}
\usepackage{graphicx}
\usepackage{subfigure}

\def\vph{\varphi}
\def\veps{\varepsilon}
\def\vrh{\varrho}

\begin{document}

\title{Anomalous Abelian Solitons}

\author{Matthias Schmid} 
\email{Matthias.Schmid@epfl.ch}
\affiliation{
  Institut de Th\'eorie des Ph\'enom\`enes Physiques,
  Ecole Polytechnique F\'ed\'erale de Lausanne,
  CH-1015 Lausanne, Switzerland}

\author{Mikhail Shaposhnikov}
\email{Mikhail.Shaposhnikov@epfl.ch}
\affiliation{
  Institut de Th\'eorie des Ph\'enom\`enes Physiques,
  Ecole Polytechnique F\'ed\'erale de Lausanne,
  CH-1015 Lausanne, Switzerland}

\date{\today}

\begin{abstract}

The chiral Abelian Higgs model contains an interesting class of solitons
found by Rubakov and Tavkhelidze. These objects carry non-zero
fermion number $N_F$ (or Chern-Simons number $N_{CS}$, what is the
same because of the chiral anomaly) and are stable for sufficiently large
$N_F$. In this paper we study the properties of these anomalous solitons.
We find that their energy-versus-fermion-number ratio is given by
$E\sim N_{CS}^{3/4}$ or $E\sim N_{CS}^{2/3}$ depending on the structure
of the scalar potential. For the former case we demonstrate that there is a
lower bound on the soliton energy, which reads  $E \geq c\ N_{CS}^{3/4}$,
where $c$ is some parameter expressed through the masses and coupling constants of the
theory. We construct the anomalous solitons numerically accounting both for
Higgs and gauge dynamics and show that they are not spherically symmetric.
The thin wall approximation valid for macroscopic solutions with $N_{CS} \gg 1$
is discussed as well.

\end{abstract}

\maketitle

\section{\label{sec:intro}Introduction}

Solitons -- stable localized solutions to the classical equations of
motion of non-linear field theory represent an interesting class of
particle-like states in quantum field theory. Well known examples
include topological solitons such as the kink in $1+1$ dimensions
\cite{Finkelstein:1959fs,Finkelstein:1966ft}, the vortex in $2+1$ dimensions
\cite{Abrikosov:1956sx,Nielsen:1973cs}, the monopole \cite{Polyakov:1974ek,'tHooft:1974qc}
and the skyrmion \cite{Skyrme:1962vh} in $3+1$ dimensions. The stability of these
solutions is ensured by topological reasons. Another class of
solutions -- non-topological solitons or Q-balls \cite{Friedberg:1976me,Coleman:1985ki},
are stable because of conservation of some global Abelian \cite{Coleman:1985ki}
or non-Abelian \cite{Safian:1987pr} charge. The above mentioned objects exist in pure
bosonic theories. Yet another type of solitons, ``electroweak bags"
\cite{Khlebnikov:1986ky,Copeland:1988bi}, use the fact that fermions
can be trapped inside a spherical cavity created by a (otherwise unstable)
space-dependent scalar field.

In \cite{Rubakov:1986am}, referred to as RT in the following, a new class of solitons was
found, for which the presence of chiral fermions and Abelian gauge symmetry play the key
role. In very general terms, the construction is based on the following
observation. Consider an Abelian Higgs model with chiral fermions and
arrange the Yukawa couplings in such a way that all fermions get a
mass $m_F$ from the Higgs condensate. Then, the energy of $N_F$
well separated fermions is $E_F \sim m_F N_F$. Due to the chiral anomaly 
these fermions can be converted to a gauge configuration carrying the
Chern-Simons number $N_{CS}= N_F$, and the energy of the gauge-Higgs
system $E$ with non-zero $N_{CS}$ may appear to be smaller than
$E_F$. If true, then one gets a stable soliton characterized by
$N_{CS}$. Indeed it was demonstrated in RT
using a variational principle that the energy $E$ of the system with 
$N_{CS}\neq 0$ is bounded from {\em above} by $ E_0 N_{CS}^{3/4}$,
where $E_0$ is some constant. Thus the bosonic configuration is always
energetically more favorable than the collection of $N_F$ fermions at
sufficiently large $N_{CS}$. The stable bosonic configuration will be
referred to as ``anomalous Abelian soliton'' in the following. In fact,
anomalous Abelian solitons have some common features with Hopf solitons, studied in
\cite{Faddeev:1996zj}. 

The Abelian character of the gauge group is essential
for absolute stability of anomalous solitons \cite{Rubakov:1985nk,Matveev:1987gq}.
Indeed, if the gauge group is non-Abelian, fermions can always be converted
into a gauge {\em vacuum} configuration \cite{'tHooft:1976up,Jackiw:1976pf,Callan:1976je},
which may carry arbitrary integer Chern-Simons number. In other words, non-Abelian anomalous
solitons, if they exist, can only be metastable.

Interestingly, anomalous solitons may potentially exist \cite{Rubakov:1985nk,Matveev:1987gq} in
the Standard Model, since it contains an Abelian U(1) gauge group and
chiral fermions. This issue, however, has not been clarified yet. 

This paper is devoted to the detailed study of anomalous Abelian
solitons. We demonstrate, with the use of different inequalities from
functional analysis, that for a certain class of scalar potentials
also a {\em lower} bound on the soliton mass,
$E \geq c\ N_{CS}^{3/4}$, can be established, where $c$ is some
parameter related to the masses and coupling constants of the underlying
theory. We construct the anomalous solitons numerically, taking into
account both the dynamics of the gauge and the Higgs field, and find that
the solitons are not spherically symmetric. We also discuss the thin wall
approximation valid in the limit of large Chern-Simons number, which will
allow us to remove the Higgs dynamics from consideration.

The organisation of the article is as follows. In the next section, we
review the Tavkhelidze and
Rubakov construction \cite{Rubakov:1985nk,Rubakov:1986am} of
anomalous solitons, in order to make the paper self-contained. In
Sec.~\ref{sec:Soliton} we derive a lower bound on the energy of the
solitons and discuss their general properties. In Sec.~\ref{sec:Numerics}
we present the numerical solutions of the field equations as a function
of Chern-Simons number and coupling constants of the theory.
Section \ref{sec:domain shape} is devoted to the solution in the thin wall
approximation. Finally, we summarize our main results in Sec.~\ref{sec:summary}.

\section{\label{sec:Model}The Rubakov-Tavkhelidze soliton}
\subsection{\label{sec:Prel}The model} 
Consider an Abelian Higgs model with the complex scalar field
$\Phi$, a pair of left-handed fermions with opposite
charges  $\psi_{L}^{1}$ and $\psi_{L}^{2}$, a pair of neutral right-handed
fermions $\psi_{R}^{1}$ and $\psi_{R}^{2}$, and chiral interaction between the
fermions and the $U(1)$ gauge field. The Lagrangian is
\begin{eqnarray}
\mathcal{L}=&&-\frac{1}{4}F_{\mu\nu}F^{\mu\nu}+\left|D_\mu\Phi\right|^2-V(|\Phi|)\nonumber\\
&&+i\bar{\psi}_{L}^{1} \gamma^\mu\left(\partial_\mu-ig A_\mu\right)\psi_{L}^{1}+i\bar{\psi}_{R}^{1}\gamma^\mu\partial_\mu\psi_{R}^{1}\nonumber\\
&&+i\bar{\psi}_{L}^{2} \gamma^\mu\left(\partial_\mu+ig A_\mu\right)\psi_{L}^{2}+i\bar{\psi}_{R}^{2}\gamma^\mu\partial_\mu\psi_{R}^{2}\nonumber\\
&&-\lambda_1 \left(\bar{\psi}_{L}^{1}\psi_{R}^{1}\Phi+\mathrm{h.c.}\right)
-\lambda_2\left(\bar{\psi}_{L}^{2}\psi_{R}^{2}\Phi^*+\mathrm{h.c.}\right)
\end{eqnarray}
with the potential
\begin{equation}\label{eq:scalar potential}
V(|\Phi|)=\lambda(|\Phi|^2-v^2)^2
\end{equation}
and $D_\mu=\partial_\mu-igA_\mu$. The vacuum expectation value (VEV) of the scalar field is equal
to $v$ and the gauge and Higgs bosons obtain the masses $m_V=\sqrt{2}gv$ and
$m_H=2\sqrt{\lambda}v$, respectively. For simplicity we choose the two Yukawa couplings to be equal
to each other, $\lambda_F\equiv\lambda_1=\lambda_2$, such that all fermions have equal masses
$m_F=\lambda_F v$. Note that we are forced to introduce at least two fermions in order to make
the theory free from gauge anomalies. 

The total fermionic current
\begin{equation*}
j_F^\mu=\sum_{i=1}^2\left(\bar{\psi}_{L}^{i}\gamma^\mu\psi_{L}^{i}+\bar{\psi}_{R}^{i}\gamma^\mu\psi_{R}^{i}\right)
\end{equation*}
is anomalous in this model,
\begin{equation}
\partial_\mu j_F^\mu=-f\frac{g^2}{32\pi^2}F_{\mu\nu}\tilde{F}^{\mu\nu}\ ,
\end{equation}
where $f=2$ is the number of left-handed fermions.

\subsection{\label{sec:instab}Instability of fermionic matter}
In this subsection we follow \cite{Rubakov:1985nk,Rubakov:1986am} in order to explain why the fermionic matter
becomes unstable at sufficiently high fermionic density.

At zero fermionic density, the ground state of the boson fields is $\bm{A}=0$ and $|\Phi|=v$,
which is called the \emph{normal} state, following RT. Consider now a homogeneous system of infinite volume
containing fermions with number density $n_F$. The system is supposed to be neutral with respect to
the gauge charge. We only consider the weak-coupling limit ($\lambda\ll 1$ and $g^2\ll 1$), where we can treat the
fields $A_\mu$ and $\Phi$ as classical condensates. The neutrality of the system implies
$A_0=0$ and that the fields $\bm{A}$ and $\Phi$ are time-independent. The fermions can be characterized by the chemical
potential $\mu_F$, which is, at zero temperature, the energy up to which the Fermi levels
are filled. The fermion number density $n_F$ is related to the Fermi energy $\mu_F$ by
\begin{equation*}
n_F=f\frac{\mu_F^3}{3\pi^2}\ .
\end{equation*}
We will use the unitary gauge $\mathrm{Im}(\Phi)=0$ and denote $\phi=\mathrm{Re}(\Phi)$ in the following.
The fermions can be integrated out \cite{Redlich:1984md} and, up to corrections of order $\mu_F^{-1}$,
the static energy functional of the bosonic fields is given by
\begin{equation}
\label{eq:eff hamiltonian}
E_B[\bm{A},\phi]=\int\left[\frac{\bm{B}^2}{2}+(\bm{\nabla}\phi)^2
+g^2\bm{A}^2\phi^2+\lambda(\phi^2-v^2)^2
-\frac{f\mu_F g^2}{32\pi^2}\veps^{ijk}F_{ij}A_k\right]\ d^3x\ ,
\end{equation}
where $\bm{B}=\bm{\nabla}\times\bm{A}$ is the magnetic field.
The first four terms in \eqref{eq:eff hamiltonian} are the classical energy density of bosons,
while the last (Chern-Simons) term is due to the interaction with fermions. The physical reason
for the appearance of the Chern-Simons term is as follows. As the gauge field
$\bm{A}$ increases, some fermionic energy levels cross the zero-energy line and the number
of real fermions decreases by $fN_{CS}$, where
\begin{equation}
\label{eq:abelian NCS}
N_{\text{CS}}[\bm{A}]=\frac{g^2}{32\pi^2}\int \veps^{ijk}F_{ij}A_k\ d^3x
=\frac{g^2}{16\pi^2}\int\bm{A}\cdot\bm{B}\ d^3x
\end{equation}
is the Chern-Simons number of the gauge field.

The quadratic part of the static energy functional \eqref{eq:eff hamiltonian} has a negative mode for
\begin{equation*}
f\mu_F>\mu_{\text{crit}}\ ,
\end{equation*}
where
\begin{equation}
\mu_{\text{crit}}=\frac{16\pi^2}{g^2}m_V\ .
\end{equation}
Thus the normal ground state of fermionic matter is absolutely unstable at
\begin{equation}
n_F>n_{\text{crit}}=\frac{\mu_{\text{crit}}^3}{3\pi^2f^2}\ .
\end{equation}
The negative mode is given by
\begin{equation}
\bm{e}(\bm{x})=\left(\bm{e}_1\cos(\bm{k}\cdot\bm{x})+\bm{e}_2\sin(\bm{k}\cdot\bm{x})\right)\ ,
\end{equation}
where $k=m_V$ and $\bm{e}_{1,2}$ are real polarization vectors
orthogonal to each other and to $\bm{k}$. Note that a perturbation $\bm{A}(\bm{x})=a\bm{e}(\bm{x})$, where $a$ is a small
amplitude has the Chern-Simons density
\begin{equation*}
n_{CS}=\frac{g^2}{16\pi^2}\bm{A}\cdot\bm{B}=\frac{g^2}{16\pi^2}ka^2\ .
\end{equation*}
As the amplitude of the unstable mode grows, the term $g^2\bm{A}^2\phi^2$ in the energy acts as a positive
mass term for the Higgs field, which leads to the disappearance of the Higgs field condensate. The
system undergoes a transition to a state with $\phi=0$ containing a negligible number of real fermions
and a gauge field condensate $\bm{A}\neq 0$ with non-zero Chern-Simons number.
We will call this state the \emph{abnormal} state, as in RT. 
\subsection{\label{sec:macro descr}Domain of abnormal matter}
Still following \cite{Rubakov:1985nk,Rubakov:1986am}, we show in this subsection, that a finite domain of the abnormal
state with sufficiently large Chern-Simons number has a lower energy than a system containing $N_F=fN_{CS}$
real fermions. The stable configuration of gauge and scalar field condensate is the minimum of the static energy functional
for the bosonic fields
\begin{equation}
\label{eq:static energy}
E[\bm{A},\phi]=\int\left[\frac{\bm{B}^2}{2}+(\bm{\nabla}\phi)^2+g^2\bm{A}^2\phi^2+\lambda(\phi^2-v^2)^2\right]\ d^3x\ ,
\end{equation}
under the constraint of a constant Chern-Simons number \eqref{eq:abelian NCS}. Varying the functional
\begin{equation}
E[\bm{A},\phi]-\mu\frac{16\pi^2}{g^2} N_{CS}[\bm{A}]
\end{equation}
with respect to $\bm{A}$ and $\phi$, where $\mu$ is a Lagrangian multiplier, we get the following field equations
\begin{subequations}
\label{eq:Euler-Lagrange}
\begin{align}
\bm{\nabla}^2\phi-g^2\bm{A}^2\phi-2\lambda(\phi^2-v^2)\phi&=0\ ,\\
\label{eq:force-free A}
\bm{\nabla}\times\left(\bm{\nabla}\times\bm{A}\right)-2\mu\left(\bm{\nabla}\times\bm{A}\right)+2g^2\phi^2\bm{A}&=0\ .
\end{align}
\end{subequations}
Solutions of Eqs.~\eqref{eq:Euler-Lagrange} with finite energy have to satisfy the boundary conditions
\begin{equation*}
\phi\rightarrow v\quad\text{and}\quad\bm{A}\rightarrow 0\quad\text{for}\quad|\bm{x}|\rightarrow\infty\ .
\end{equation*}
For a static solution $\{\bm{A}_{cl}(\bm{x}),\phi_{cl}(\bm{x})\}$ of Eqs.~\eqref{eq:Euler-Lagrange} we obtain
\begin{equation}
\label{eq:classical static energy}
E[\bm{A}_{cl},\phi_{cl}]=\mu\frac{16\pi^2}{g^2}N_{CS}+\int\left[(\nabla\phi_{cl})^2+\lambda(\phi_{cl}^2-v^2)^2\right]\ d^3x\ ,
\end{equation}
where the first term in Eq.~\eqref{eq:static energy} has been integrated by parts and \eqref{eq:force-free A} has been used.

Let us consider a compact domain of finite volume $V$ in the abnormal state, embedded into the normal vacuum. We suppose
that the size $R\sim V^{1/3}$ of the domain is much larger than the two characteristic length scales
in the model, the thickness of the domain wall $\sim 1/m_H$ and the penetration length $\sim 1/m_V$ of the magnetic field.
Then, the contribution of the scalar field to the energy (the second term of Eq.~\eqref{eq:classical static energy})
can be written
\begin{equation}
E_{\text{scal}}=\int\left[(\nabla\phi_{cl})^2+\lambda(\phi_{cl}^2-v^2)^2\right]\ d^3x\approx\lambda v^4 R^3\ I_V\ ,
\end{equation}
where $I_V=V/R^3$ is the shape factor for the domain $V$. The surface energy, which is of order $R^2 v^3$,
can be neglected. Since we want to minimize the total energy,
\begin{equation}
E=\mu\frac{16\pi^2}{g^2}N_{CS}+\lambda v^4 R^3\ I_V\ ,
\end{equation}
with respect to $R$, we have to find how the Lagrangian multiplier $\mu$ scales with $R$. Since $\phi=0$ inside the domain
of the abnormal state, it follows from Eq.~\eqref{eq:force-free A} that $\mu\sim 1/R$.
Thus we can write $\mu=\nu/(2R)$, where $\nu$ is a numerical coefficient. The
total energy as a function of $R$ becomes
\begin{equation}
\label{eq:E(R)}
E(R)=8\pi^2\frac{\nu}{g^2}\frac{N_{CS}}{R}+\lambda v^4 R^3 I_V
\end{equation}
and minimizing $E(R)$ with respect to $R$ leads to
\begin{eqnarray}
\label{eq:domain size}
R&=&\left(\frac{64\pi^2\nu}{3I_V}\right)^{1/4}\frac{1}{m_V}\left(\frac{N_{CS}}{\beta}\right)^{1/4}\ ,\\
\label{eq:total energy}
E&=&\sqrt{2}\left(\frac{16\pi^2\nu}{3}\right)^{3/4}\left(I_V\beta\right)^{1/4}\frac{m_V}{g^2}N_{CS}^{3/4}\ ,
\end{eqnarray}
where we expressed $R$ and $E$ in terms of the vector boson mass $m_V$ and the parameter $\beta$ defined by
\begin{equation}
\label{eq:beta parameter}
\beta=\frac{m_H^2}{m_V^2}=\frac{2\lambda}{g^2}\ .
\end{equation}
Both the size of the domain $R$ and the total energy $E$ grow slowly with increasing $N_{CS}$. If $N_{CS}>N_{\text{crit}}$,
the energy \eqref{eq:total energy} is smaller than the energy of the normal state with $N_F=fN_{CS}$ real fermions for which $E_F\sim m_F N_F$.
Parametrically, $N_{\text{crit}}\sim (\frac{m_V}{g^2 m_F})^4$. Therefore, the domain of abnormal matter
is stable. Since $R\sim v^{-1}N_{CS}^{1/4}$, we get $R\gg v^{-1}$ for large $N_{CS}$, and
the surface energy $R^2 v^3\sim v N_{CS}^{1/2}$ becomes indeed negligible at large $N_{CS}$, as we assumed above.

In \cite{Rubakov:1985nk} the domain was considered to be a sphere of radius $R$ and
Eq.~\eqref{eq:force-free A} was solved analytically (remember that $\phi\equiv 0$ inside the domain).
The solution can be expressed in terms of Bessel functions and one gets
\begin{equation*}
\mu=\frac{\xi_0}{2R}\ ,
\end{equation*}
where $\xi_0$ is the first node of the Bessel function $J_{3/2}$. Inserting $\nu=\xi_0$ and $I_V=4\pi/3$
into Eqs.~\eqref{eq:domain size} and \eqref{eq:total energy}
we find exactly Eqs.~(6.7) and (6.8) of \cite{Rubakov:1985nk}.

This completes our review of the main results in RT, who proved the existence of anomalous
Abelian solitons. The remainder of the present paper is devoted to the detailed study of the structure of these solitons.
In particular, we would like to answer the following questions:
\begin{itemize}
\item
What is the structure of the gauge and Higgs field condensates of anomalous solitons?
\item
What is their exact shape? Are they spherically symmetric?
\item
Is the exponent $3/4$ in the power-law dependence of the energy on $N_{CS}$ universal for anomalous solitons?
\item
Does the structure of the solitons depend on the choice of the scalar potential?
\item
What happens at small $N_{CS}$, when the dynamics of the scalar field has to be taken into account?
\end{itemize}
\section{\label{sec:Soliton} Properties of the soliton}
\subsection{\label{sec:lower bound}Lower bound on the energy}
In the following we derive a lower bound on the energy of anomalous solitons, which reads
\begin{equation}
\label{eq:ineqVak}
E\geq c N_{CS}^{3/4}\ ,
\end{equation}
where $c$ is a constant depending on the VEV of the Higgs field and the coupling constants. Such a bound on the
energy is rather unusual because of the fractional power of the topological charge. A bound of this type was first
obtained by Vakulenko and Kapitanksy \cite{Vakulenko:1979uw} for solitons of the non-linear $\sigma$-model, for which
the topological charge is the Hopf number. The Vakulenko bound was later improved (i.e.~increasing the constant $c$) by
Kundu and Rybakov \cite{Kundu:1982bc} and Ward \cite{Ward:1998pj}.

The derivation of \eqref{eq:ineqVak} is based on the use of some inequalities of functional analysis, which we
review in appendix \ref{App:A}. Let us start with an estimate of the
Chern-Simons number \eqref{eq:abelian NCS},
\begin{eqnarray}
\label{eq:Holder}
N_{CS}&=&g^2c_1\int\bm{A}\cdot\bm{B}\ d^3x\nonumber\\
&\leq& g^2c_1\left(\int |\bm{B}|^{6/5}\ d^3x\right)^{5/6}\left(\int |\bm{A}|^6\ d^3x\right)^{1/6}\ ,
\end{eqnarray}
where $c_1=1/(16\pi^2)$ and we have used the H\"older inequality \eqref{eq:Holder inequality for vectors} for $p=6/5$ and $q=6$. The first
factor on the r.h.s. of \eqref{eq:Holder} is estimated using \eqref{eq:modified Hölder},
\begin{equation}
\label{eq:modified Holder inequality}
\left(\int |\bm{B}|^{6/5}\ d^3x\right)^{5/6}\leq \left(\int |\bm{B}|^2\ d^3x\right)^{1/6}\left(\int |\bm{B}|\ d^3x\right)^{2/3}\ ,
\end{equation}
and the second factor with the use of the Gagliardo-Nirenberg-Sobolev inequality \eqref{eq:Sobolev inequality},
\begin{equation}
\label{eq:GNS}
\left(\int |\bm{A}|^6\ d^3x\right)^{1/6}\leq c_2\left(\int\bigl|\nabla|\bm{A}|\bigr|^2\ d^3x\right)^{1/2}\ .
\end{equation}
Rosen \cite{rosen:30} found the smallest possible value for the constant $c_2$ in the previous inequality to be
\begin{equation*}
c_2=\frac{1}{\sqrt{3}}\left(\frac{2}{\pi}\right)^{2/3}\ .
\end{equation*}
The integrand of \eqref{eq:GNS} is further bounded from above,
\begin{eqnarray}
\label{eq:coulomb gauge}
\bigl|\nabla|\bm{A}|\bigr|^2&=&\frac{1}{|\bm{A}|^2}\sum_{i=1}^{3}\left(\bm{A}\cdot\partial_i\bm{A}\right)^2\leq\sum_{i,j=1}^{3}(\partial_i A_j)^2\nonumber\\
&=&|\bm{\nabla}\times\bm{A}|^2+\sum_{i,j=1}^{3}\partial_i A_j \partial_j A_i\ .
\end{eqnarray}
Imposing the Coulomb gauge $\bm{\nabla}\cdot\bm{A}=0$, the last term in Eq.~\eqref{eq:coulomb gauge} vanishes through
integration by parts and we obtain
\begin{equation}\label{eq:gradient estimate}
\int\bigl|\nabla|\bm{A}|\bigr|^2\ d^3x\leq\int |\bm{B}|^{2}\ d^3x\ .
\end{equation}
Inserting \eqref{eq:modified Holder inequality}, \eqref{eq:GNS} and \eqref{eq:gradient estimate}
into \eqref{eq:Holder} yields
\begin{equation}
N_{CS}\leq g^2c_1 c_2\left(\int |\bm{B}|^{2}\ d^3x\right)^{2/3}\left(\int |\bm{B}|\ d^3x\right)^{2/3}\ ,
\end{equation}
or equivalently,
\begin{equation}
\label{eq:ineq with C}
N_{CS}^{3/2}\leq g^3 C\left(\int|\bm{B}|^{2}\ d^3x\right)\left(\int|\bm{B}|\ d^3x\right)\ ,
\end{equation}
where
\begin{equation}
\label{eq:constant Rosen}
C=(c_1 c_2)^{3/2}=\frac{1}{32\pi^4 3^{3/4}}\ .
\end{equation}
At this point, it is worth mentioning \cite{Ward:1998pj}, where it was claimed that the constant $C$ can even be
reduced to $C=1/(256\pi^4)$. If we define an average magnetic field by
\begin{equation}
\bar{B}=\frac{\int |\bm{B}|^{2}\ d^3x}{\int |\bm{B}|\ d^3x}\ ,
\end{equation}
we can write down an inequality for the magnetic energy
\begin{equation}
\label{eq:ineqEmagn}
E_{\text{magn}}=\int\frac{|\bm{B}|^{2}}{2}\ d^3x\geq \frac{1}{2}\left(\frac{\bar{B}}{C}\right)^{1/2}\left(\frac{N_{CS}}{g^2}\right)^{3/4}\ .
\end{equation}

In order to obtain an inequality for the total energy rather than for $E_{\text{magn}}$, we consider the scale transformation 
\begin{align*}
\phi_\Lambda(\bm{x})&=\phi_{cl}(\Lambda\bm{x})\ ,\\
\bm{A}_\Lambda(\bm{x})&=\Lambda\bm{A}_{cl}(\Lambda\bm{x})\ ,
\end{align*}
where $\{\phi_{cl}(\bm{x}),\ \bm{A}_{cl}(\bm{x})\}$ is a static solution of the field equations with finite energy. This transformation leaves
the Chern-Simons number unchanged. The static energy
\begin{equation*}
E(\Lambda)=E[\bm{A}_\Lambda(\bm{x}),\phi_\Lambda(\bm{x})]
\end{equation*}
must have a minimum at $\Lambda=1$. We get for the energy
\begin{equation*} 
E(\Lambda)=\Lambda\int \frac{\bm{B}^2_{cl}}{2}\ d^3y+\frac{1}{\Lambda}\int\left[(\bm{\nabla}\phi_{cl})^2+g^2\bm{A}_{cl}^2\phi_{cl}^2\right]\ d^3y
+\frac{1}{\Lambda^3}\int V(\phi_{cl})\ d^3y\ ,
\end{equation*}
where $\bm{y}=\Lambda\bm{x}$. Requiring that
\begin{equation*}
\frac{\partial E}{\partial\Lambda}\Big|_{\Lambda=1}=0
\end{equation*}
leads to the relation
\begin{equation}\label{eq:Derricks relation}
\int\frac{\bm{B}_{cl}^2}{2}\ d^3y=\int\left[(\bm{\nabla}\phi_{cl})^2+g^2\bm{A}_{cl}^2\phi_{cl}^2+3V(\phi_{cl})\right]\ d^3y\ ,
\end{equation}
which is exact and has to be satisfied by any static solution of the field
equations. From \eqref{eq:Derricks relation} we deduce
\begin{equation}
E=\frac{4}{3}E_{\text{magn}}+\frac{2}{3}\int\left[(\bm{\nabla}\phi_{cl})^2+g^2\bm{A}_{cl}^2\phi_{cl}^2\right]\ d^3x
\end{equation}
and using \eqref{eq:ineqEmagn} we get
\begin{equation}
E\geq\frac{4}{3}E_{\text{magn}}\geq\frac{2}{3}\left(\frac{\bar{B}}{C}\right)^{1/2}\left(\frac{N_{CS}}{g^2}\right)^{3/4}\ ,
\end{equation}
which is a lower bound for the energy of anomalous solitons, provided the
average magnetic field $\bar{B}$ is bounded from below. We will derive such a
bound in the next subsection using known results from Ginzburg-Landau theory of superconductors.
\subsection{\label{sec:superconductors}Anomalous solitons and superconductors}
The static energy \eqref{eq:static energy} is just the relativistic version of
the Ginzburg-Landau free energy for a superconductor (see e.g.~\cite{Landau:vol9}).
The state with $\phi=v$ corresponds to the superconducting state, while $\phi=0$
coincides with the normal-conducting state. It is well known from
superconductivity theory that the superconducting state is destroyed by large
external magnetic fields and that the way in which this superconductivity breaking
occurs, depends on the ratio $\beta$ of the Higgs
mass to the gauge boson mass squared (cf. Eq.~\eqref{eq:beta parameter}).

In superconductors of the first kind $(\beta<1)$ the superconducting state can persist
only up to a critical magnetic field (see e.g.~\cite{1960ecm}), which is given
by $B_c=(2\lambda)^{1/2}v^2$. At $B>B_c$ an intermediate state is formed, in which
normal-conducting domains (with $\phi=0$ and $B\geq B_c$) emerge within the
superconducting state ($\phi=v$, $B=0$). Thus for type I superconductors the lower bound
on the average magnetic field is provided by the critical magnetic field, i.e.
\begin{equation*}
\bar{B}\geq B_c=\frac{\sqrt{\beta}}{2g} m_V^2
\end{equation*}
and we finally get the lower bound
\begin{equation}
\label{eq:final bound}
E\geq \frac{\sqrt{2}}{3}C^{-1/2}\beta^{1/4}\frac{m_V}{g^2}N_{CS}^{3/4}\quad\text{for}\quad\beta\leq 1\ .
\end{equation}

On the other hand, for superconductors of the second kind ($\beta>1$) the magnetic field penetrates
the superconductor along vortex filaments with quantized magnetic flux (Abrikosov
vortices~\cite{Abrikosov:1956sx}). For $B_{c1}<B<B_{c2}$ the superconducting
sample is in a mixed state forming a lattice of Abrikosov vortices. The lower critical
field $B_{c1}$ marks the point, when the first vortex filaments appears and is given by
\begin{equation*}
B_{c1}=\frac{\veps}{\Phi_0}\ ,
\end{equation*}
where $\veps$ is the energy per unit length of one vortex filament and $\Phi_0$ is the unit
flux quantum. For $\beta\gg 1$ the vortex energy can be calculated analytically and the lower
critical field becomes~\cite{Landau:vol9}
\begin{equation}
B_{c1}=\frac{B_c}{2\sqrt{\beta}}\ln{\left(\frac{\beta}{2}\right)}\ .
\end{equation}
If we suppose, that for $\beta\gg 1$ the state with the lowest energy is a
lattice of vortices of total length $L$, we can write the average magnetic field as
\begin{equation}
\bar{B}=\frac{\int |\bm{B}|^{2}\ d^3x}{\int |\bm{B}|\ d^3x}\sim\frac{2L\veps}{L\Phi_0}=2B_{c1}\ ,
\end{equation}
and we obtain for the lower bound
\begin{equation}
\label{eq:final bound type 2}
E\geq \frac{\sqrt{2}}{3}C^{-1/2}\left[\ln{\left(\frac{\beta}{2}\right)}\right]^{1/2}\frac{m_V}{g^2}N_{CS}^{3/4}
\quad\text{for}\quad\beta\gg 1\ .
\end{equation}

\subsection{\label{sec:shape} Shape of the solitons}
From (macroscopic) superconductivity theory we can also gain an insight on the shape of anomalous solitons
at large $N_{CS}$. Consider a surface separating a normal-conducting and a superconducting domain
within a type I superconductor in the intermediate state. Because in the superconducting domain $\bm{B}=0$, the
continuity of the orthogonal component $B_\perp$ across the surface tells us, that the magnetic
field in the normal-conducting domain has to be tangential on this surface. Furthermore the magnetic field
has to be equal to the critical field $B_c$. Therefore we expect, at least for $\beta<1$, the magnetic field
to be tangential and equal to $B_c$ on the boundary of the solitons. However, it is not possible to contruct
a constant tangent vector field on a sphere. More generally, on a closed surface of genus 0 there is no continuous tangent
vector field, which is non-zero everywhere on that surface (Poincar\'e-Hopf theorem). The simplest manifold on which such a vector field can
exist, is a torus (the surface has to be of genus 1). Consequently, the highest possible symmetry of anomalous solitons
is axial symmetry.

\subsection{\label{sec:symmPot energy} Dependence on $V(\phi)$}
Up to this point we have discussed anomalous solitons for the potential \eqref{eq:scalar potential}.
In this subsection we consider solitons for different scalar potentials. We shall
see, that the $3/4$ power-law of the soliton energy, $E\sim N_{CS}^{3/4}$, is not universal, but depends
on the scalar potential.

\begin{figure}[tbh]
\subfigure{\label{fig:HiggsPot}\includegraphics{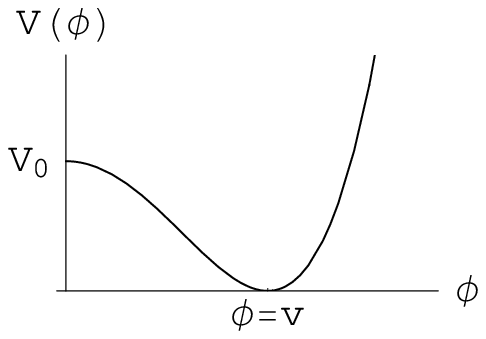}}
\subfigure{\label{fig:symmPot}\includegraphics{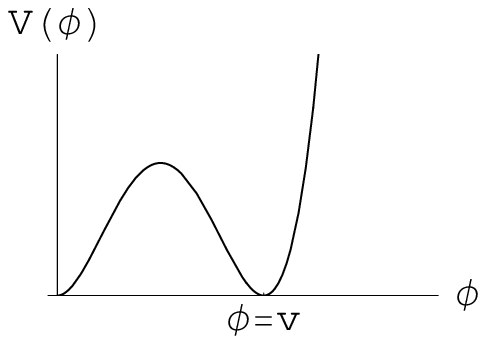}}
\subfigure{\label{fig:unstPot}\includegraphics{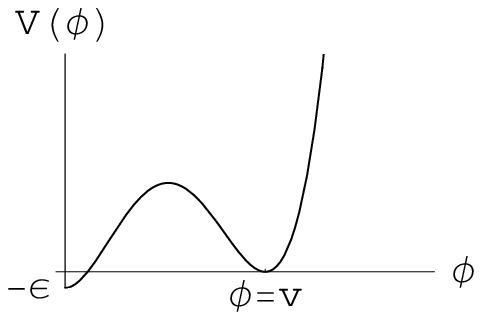}}
\caption{Three choices for the scalar potential $V(\phi)$: (a) Potential with $V(\phi=0)=V_0>0$ (non-zero
critical magnetic field). (b) Potential with $V(\phi=0)=0$ (zero critical magnetic field).
(c) Potential with $V(\phi=0)<0$ (the ground state $\phi=v$ is metastable).}
\label{fig:potentials}
\end{figure}

Let us consider three different classes of potentials shown in Fig.~\ref{fig:potentials}. For
%
potentials with $V(\phi=0)=V_0>0$, such as the potential \eqref{eq:scalar potential} (see
Fig.~\ref{fig:HiggsPot}), there is a non-zero critical magnetic field, which is given by
\begin{equation}
\frac{B_c^2}{2}=V_0>0\ .
\end{equation}
The critical magnetic field is the energy density of the scalar field in the abnormal
state $\phi=0$. On the boundary of the two phases, $\phi=v$ and $\phi=0$, there is a pressure
balance between the pressure of the magnetic field and the vacuum energy density. The surface
contribution to the energy of the scalar field is negligible.

On the other hand for a potential with $V_0=0$ (Fig.~\ref{fig:symmPot}), the critical magnetic
field vanishes. In this case the energy of the Higgs field is dominated by a surface term
and is given by
\begin{equation}
E_{\text{scal}}=R^2 v^3 I_A\ ,
\end{equation}
where $I_A$ is a numerical coefficient. Instead of Eq.~\eqref{eq:E(R)} we obtain
\begin{equation}
E(R)=8\pi^2\frac{\nu}{g^2}\frac{N_{CS}}{R}+I_A v^3 R^2\ .
\end{equation}
Minimization with respect to $R$ shows that
\begin{eqnarray}
R&=&\left(128\pi^4\right)^{1/6}\left(\frac{g\nu}{I_A}\right)^{1/3}\frac{1}{m_V}N_{CS}^{1/3}\ ,\\
E&=&3(32\pi^8)^{1/6}\left(\frac{I_A\nu^2}{g}\right)^{1/3}\frac{m_V}{g^2}N_{CS}^{2/3}\ .
\end{eqnarray}
In the next section, we will numerically verify this result for a specific potential of the type of Fig.~\ref{fig:symmPot}.

Finally, for a potential with $V_0<0$ (see Fig.~\ref{fig:unstPot}), there is no
critical magnetic field. The normal ground state $\phi=v$ is metastable and anomalous solitons do not exist
in this case.

\section{\label{sec:Numerics}Structure of Anomalous Solitons}
In this section we construct anomalous solitons numerically. We present numerical
solutions to the field equations as a function of
the Chern-Simons number $N_{CS}$ and the parameter $\beta$. The structure of
the Higgs and gauge field condensates is analyzed in detail. Finally we
discuss the dependence of the solutions on the scalar potential.

\subsection{\label{sec:Numerical Construction}Numerical Construction}

In numerical calculations it is convenient to work in dimensionless coordinates $\widehat{\bm{x}}=m_V\bm{x}$,
and with dimensionless fields $\widehat{\phi}=\phi/v$ and $\widehat{\bm{A}}=\bm{A}/v$. In terms of these quantities
the field equations \eqref{eq:Euler-Lagrange} are
\begin{subequations}
\label{eq:hat}
\begin{eqnarray}
\widehat{\bm{\nabla}}^2\widehat{\phi}-\frac{1}{2}\widehat{\bm{A}}^2\widehat{\phi}
-\frac{\beta}{2}\left(\widehat{\phi}^2-1\right)\widehat{\phi}&=0\ ,\\
\widehat{\bm{\nabla}}\times\left(\widehat{\bm{\nabla}}\times\widehat{\bm{A}}\right)
-2\widehat{\mu}\ \widehat{\bm{\nabla}}\times\widehat{\bm{A}}+\widehat{\phi}^2\widehat{\bm{A}}&=0\ ,
\end{eqnarray}
\end{subequations}
where $\widehat{\mu}$ is the (dimensionless) Lagrangian multiplier. We want to solve
Eqs.~\eqref{eq:hat} subject to the constraint of constant Chern-Simons number,
\begin{equation}\label{eq:dimless NCS}
N_{CS}[\widehat{\bm{A}}]=\frac{1}{32\pi^2}\int\widehat{\bm{A}}\cdot\widehat{\bm{B}}\ d^3\widehat{x}\ ,
\end{equation}
which determines the Lagrangian multiplier $\widehat{\mu}$.

It followed from the discussion in section~\ref{sec:shape} that the highest possible symmetry for 
the solitons is axial symmetry. Therefore we restrict ourselves to axially
symmetric solutions and use cylindrical coordinates $(r,\vph,z)$,
for which Eqs.~\eqref{eq:hat} are reduced to a system of non-linear PDEs in the $rz$-plane:
\begin{subequations}
\label{eq:sys incl phi}
\begin{eqnarray}
\label{eq:sys ph}
\frac{1}{r}\ \partial_r\left(r\ \partial_r\phi\right)+\partial_z^2\phi
-\frac{1}{2}\left(A_r^2+A_\vph^2+A_z^2\right)\phi-\frac{\beta}{2}\left(\phi^2-1\right)\phi&=0\ ,\\
\label{eq:sys ar}
\partial_z\left(\partial_z A_r-\partial_r A_z\right)-2\mu\ \partial_z A_\vph-\phi^2 A_r&=0\ ,\\
\label{eq:sys aph}
\partial_r\left[\frac{1}{r}\ \partial_r\left(rA_\vph\right)\right]+\partial^2_z A_\vph+2\mu\left(\partial_z A_r-\partial_r A_z\right)-\phi^2 A_\vph&=0\ ,\\
\label{eq:sys az}
\frac{1}{r}\ \partial_r\left[r\left( \partial_r A_z-\partial_zA_r\right)\right]
+\frac{2\mu}{r}\left[\partial_r \left(rA_{\vph}\right)\right]
-\phi^2 A_z&=0\ .
\end{eqnarray}
\end{subequations}
For axial symmetry we have
\begin{equation}
\label{eq:sys constr}
N_{CS}=\frac{1}{8\pi}\int r A_\vph\left(\partial_z A_r-\partial_r A_z\right)drdz
\end{equation}
after integration by parts and over the angle $\vph$. Note that we are omitting hats from now on.

Solutions of Eqs.~\eqref{eq:sys incl phi} are
calculated by discretizing the fields $\phi(r,z)$ and $\bm{A}(r,z)$
on a rectangular box $[0,L]\times[-Z,Z]$ in the $rz$-plane and then solving the corresponding system of non-linear
equations using Newton's algorithm. The boundary conditions on the edges of the box are
\begin{subequations}
\label{eq:sys boundary conditions}
\begin{equation}
\phi\Big|_{r=L}=1\qquad\text{and}\qquad\phi\Big|_{z=-Z}=\phi\Big|_{z=-Z}=1
\end{equation}
for the Higgs field and
\begin{equation}
\bm{A}\Big|_{r=L}=0\qquad\text{and}\qquad\bm{A}\Big|_{z=-Z}=\bm{A}\Big|_{z=Z}=0
\end{equation}
for the gauge field. On the axis of symmetry we imposed the regularity conditions
\begin{equation}
\label{eq:reg cond}
\partial_r\phi\Big|_{r=0}=\partial_r A_z\Big|_{r=0}=0\qquad\text{and}\qquad A_r\Big|_{r=0}=A_\vph\Big|_{r=0}=0\ .
\end{equation}
\end{subequations}
For a detailed description of the numerical procedure we used to solve the PDEs, we refer the reader to appendix \ref{App B}.

\subsection{\label{sec:dep on NCS and beta}Dependence of the soliton energy on $N_{CS}$ and $\beta$}
\begin{figure}[tbh]
\includegraphics{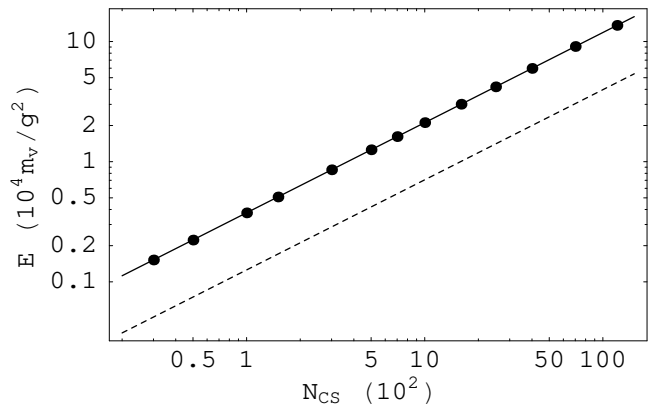}
\caption{\label{fig:EtotNCS} Logarithmic plot of the soliton energy as a function of $N_{CS}$ for $\beta=1$.
The dots show the energies calculated from the numerical solutions for $N_{CS}=30$ to $N_{CS}=12'000$.
The solid line is a fit of the form $E=a N_{CS}^{3/4}$ with $a\approx 118.826\ m_V/g^2$. The dashed line corresponds
to the lower bound \eqref{eq:final bound}.}
\end{figure}

We calculated solutions for $1<N_{CS}<1.2\cdot 10^4$ and $0.4<\beta<16$. In Fig.~\ref{fig:EtotNCS}
%
we show the energies obtained from the numerical solutions as a function of
$N_{CS}$ for $\beta=1$. The energy satisfies $E=aN_{CS}^{3/4}$, with $a\approx 118.83\ m_V/g^2$, which
confirms the $3/4$ power-law dependence on $N_{CS}$. The numerical solution gives an energy about three
times larger than the lower bound~\eqref{eq:final bound}, which is $\approx 39.7\ m_V/g^2\ N_{CS}^{3/4}$
for $\beta=1$.

\begin{figure}[tbh]
\includegraphics{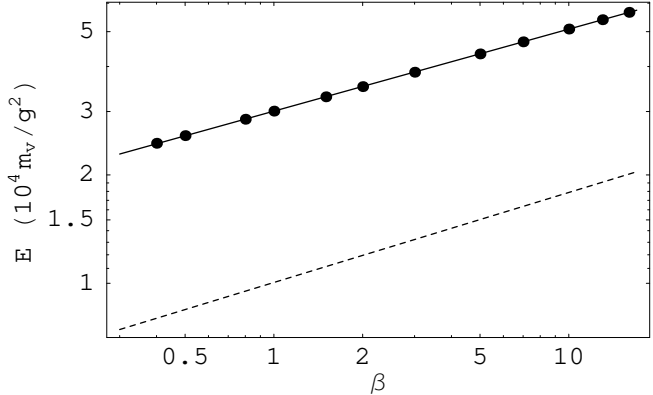}
\caption{\label{fig:EbetaNCS} Logarithmic plot of the energy as a function of $\beta$. The dots represent the energies of the
numerical solutions for $\beta$ between $0.4$ and $16$ (for $N_{CS}=1'600$). The solid line is
the fit $E=b\beta^{p}$
giving $p\approx 0.23$ and $b\approx 3.007\cdot 10^4\ m_V/g^2$. The dashed line corresponds
to the lower bound \eqref{eq:final bound}.}
\end{figure}

Fig.~\ref{fig:EbetaNCS}
%
shows the soliton energy as function of $\beta$ at $N_{CS}=1'600$.
We find $E\approx 3.01\cdot 10^4\ m_V/g^2\ \beta^p$, with $p\approx 0.23$. The
$\beta^{1/4}$ dependence is expected from Eq.~\eqref{eq:total energy}
(or from \eqref{eq:final bound}) for $\beta\leq 1$.

All the solutions we found consist of a single
domain with $\phi\ll v$ and $|\bm{B}|>B_c$, surrounded by normal vacuum. For small Chern-Simons number $N_{CS}<20$
the soliton has a size of the order of the thickness of the domain wall ($\sim 1/m_H$), but
becomes larger when $N_{CS}$ is increased.

This is the behavior we expect for $\beta\leq 1$ (cf. intermediate state of type I superconductors). 
For $\beta>1$ one could imagine solutions containing a network of closed vortices similar to
Abrikosov vortices in type II superconductors, which could have a lower energy than our solutions for
$\beta>1$. Therefore we point out, that our solutions for $\beta>1$ might be metastable with respect
to the decay into a vortex-type solution of equal Chern-Simons number. However, we did not find
any solutions of this type.

\subsection{Structure of the Higgs and the gauge fields}
Let us now look at the scalar and gauge field configurations of the solitons. We will concentrate on
the specific example with $\beta=1$ and $N_{CS}=12'000$.

\begin{figure}[tbh]
\includegraphics{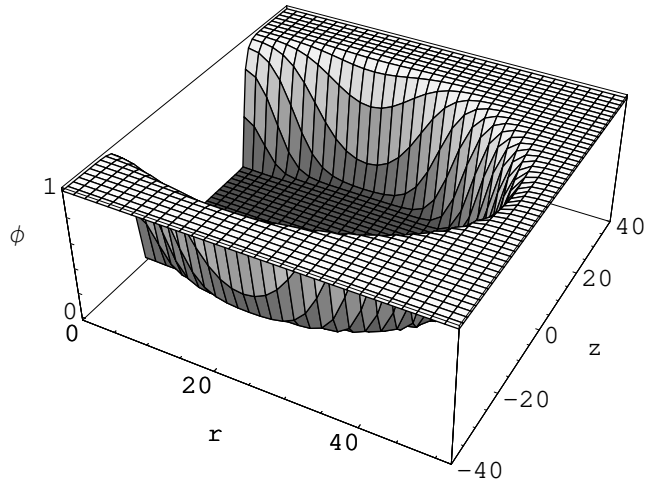}
\caption{\label{fig:Phi3D} The scalar field $\phi$ in the $rz$-plane for $N_{CS}=12'000$ and $\beta=1$.
Note that $\phi$ is shown in units of $v$ and the
coordinates $r$ and $z$ are in units of $1/m_V$. The rectangular box in the $rz$-plane used to obtain the solution
in this case was $[0,52]\times[-40,40]$.}
\end{figure}

\begin{figure}[tbh]
\includegraphics{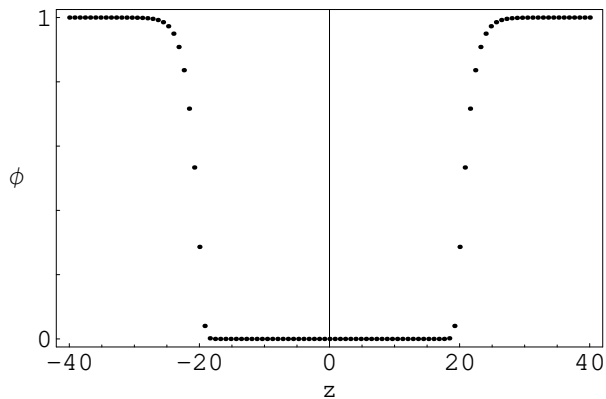}
\caption{\label{fig:Phi2D} The scalar field $\phi$ along the axis of symmetry $r=0$ (again for $N_{CS}=12'000$ and $\beta=1$).
The dots show the values of $\phi$ on the grid points of the numerical solution. The resolution is $0.8/m_V$.}
\end{figure}

Figs.~\ref{fig:Phi3D} and \ref{fig:Phi2D}
%
show a 3D-plot of the scalar field $\phi$ in the $rz$-plane and a 2D-plot of $\phi$ along
the axis of symmetry $r=0$, respectively. The thickness of the domain
wall ($\sim 1/m_H$), which separates the two phases $\phi\ll v$ and $\phi=v$,
is small compared to the size of the soliton (thin wall approximation).

\begin{figure}[tbh]
\includegraphics{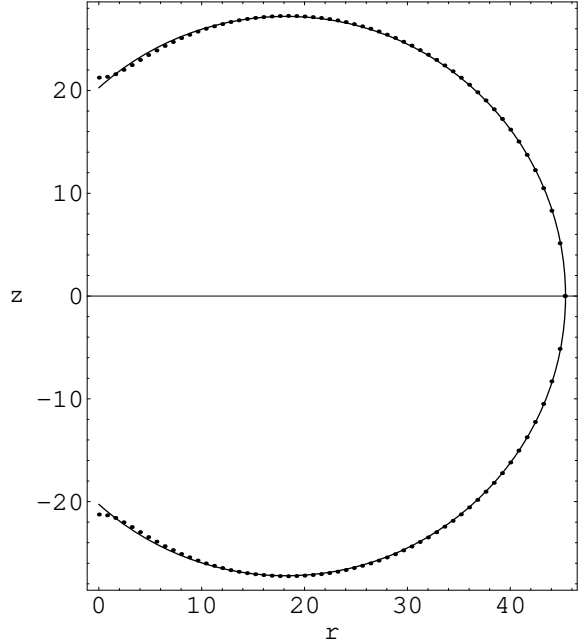}
\caption{\label{fig:SpindleFit} The dotted line shows the contour of constant scalar field $\phi=0.5$ in the $rz$-plane.
The solid line is the best fitting circle to the dotted line, which is the circle around ($18.163,0$) with radius
$r_s\approx 27.132$ (in units of $1/m_V$).}
\end{figure}

Moreover Fig.~\ref{fig:Phi3D} shows that the soliton is \emph{not} spherically symmetric.
We can determine the shape of the soliton numerically. For this end,
we define the contour of constant scalar field $\phi=0.5\ v$ as the boundary
of the soliton and fit a 2-dimensional surface to that boundary.
This is done in Fig.~\ref{fig:SpindleFit},
%
in which we show the contour $\phi=0.5\ v$ in the $rz$-plane. The contour follows very precisely a circle
of radius $r_s\approx 27.132/m_V$ centered at $(18.163/m_V,0)$. The contour deviates from a perfect circle
just in a narrow region around the axis of symmetry, whose size is of the order of a few $1/m_V$. Only in this
small region, the fields behave microscopically.
Motivated by the amazing numerical result of Fig.~\ref{fig:SpindleFit}, we conjecture that asymptotically for large
$N_{CS}$, the soliton has the shape of a so-called \emph{spindle torus}. A spindle torus is the surface of
revolution obtained by rotating a circular arc around an axis, where the arc is intersecting that axis.
However, we were not able to prove this result analytically.

\begin{figure}[tbh]
\includegraphics{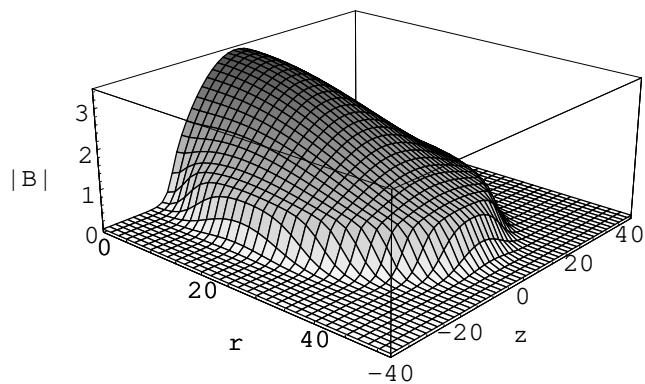}
\caption{\label{fig:BAmp3D} 3D-plot of the magnetic field $|\bm{B}|$ in the $rz$-plane in units of the
critical magnetic field $B_c$.}
\end{figure}

We now turn to the gauge field. As depicted in Fig.~\ref{fig:BAmp3D},
%
the magnetic field is confined to the region where the scalar field is vanishing or small. Inside
this domain the amplitude of the magnetic field $|\bm{B}|$ is larger than the critical magnetic field $B_c$.

\begin{figure}[tbh]
\includegraphics{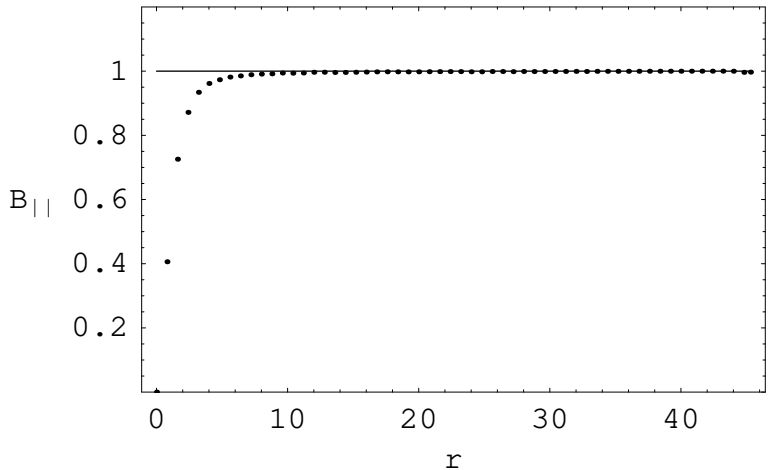}
\caption{\label{fig:Btang} The dots show the tangential component $\bm{B}_\parallel$ of the magnetic field along the boundary of the
domain as a function of the distance $r$ from the axis of symmetry (units for $\bm{B}_\parallel$: $B_c=1$). Sufficiently
far away from the axis of symmetry the magnetic field is constant and equal to the critical field $B_c$, which is represented
by the solid line.}
\end{figure}

On the boundary $|\bm{B}|=B_c$ and $\bm{B}$ is tangential to the domain boundary
(see Fig.~\ref{fig:Btang}).
%
Remember that this behavior is expected from (type I) superconductors in the intermediate state: on the 
boundaries of normal-conducting regions the magnetic field is tangential and equal to $B_c$. Outside this
boundary, where the scalar field
starts to deviate from zero, the magnetic field decays exponentially, because in the broken phase the
gauge boson becomes massive.

\begin{figure}[tbh]
\includegraphics{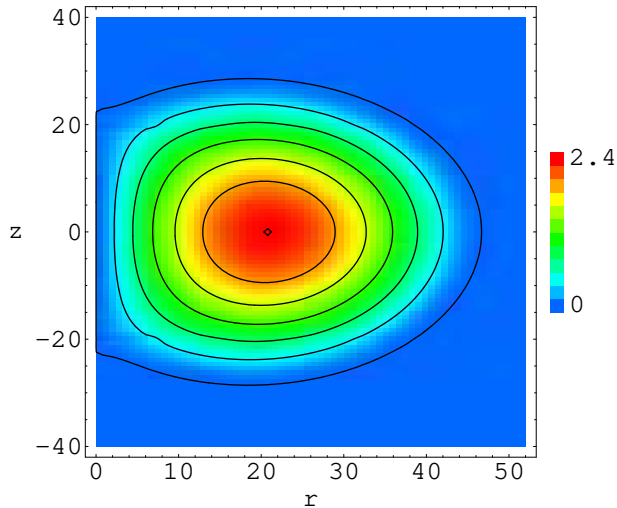}
\caption{\label{fig:Bphcont} Contour plot of the toroidal component $B_\vph$ of the magnetic field
(in units of the critical magnetic field $B_c$). The shown contours are from $0$ to $2.4$ with a step
size of $0.4$ between the contours. The maximal value of $B_\vph$ is located at 
$(r,z)\approx (20.8,0)$ and is equal to $2.4\ B_c$.}
\end{figure}

\begin{figure}[tbh]
\includegraphics{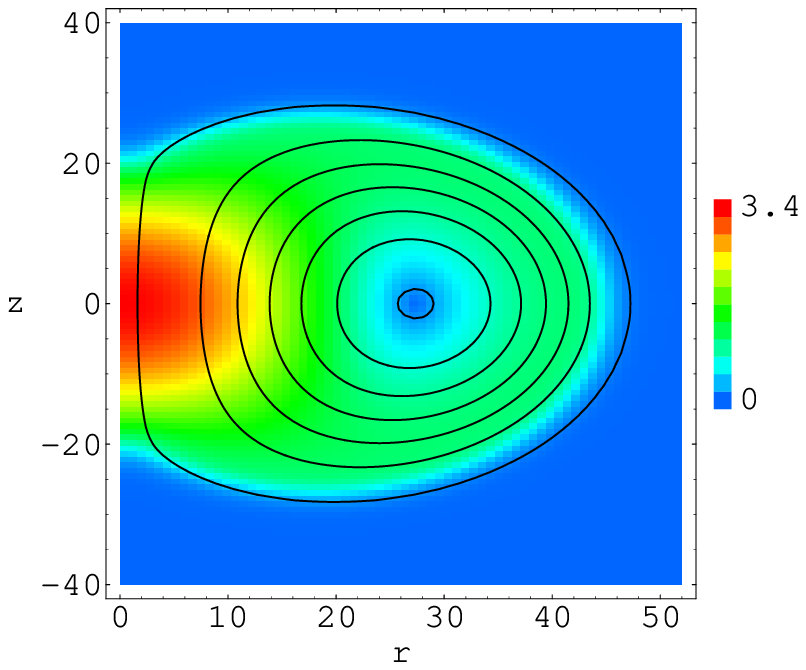}
\caption{\label{fig:Bpol} The poloidal component $\bm{B}_p$ of the magnetic field. The solid lines are projections of the field
lines onto the $rz$-plane. The colors indicate the modulus of $\bm{B}_p$ (in units of $B_c$). The poloidal field is
maximal at the origin and vanishes at $(27.4,0)$. On the boundary of the spindle torus $|\bm{B}_p|=B_c$.}
\end{figure}

To get an idea of how the magnetic field lines behave, we show in Fig.~\ref{fig:Bphcont} a
%
plot of the ``toroidal'' component $B_\vph$ and in
Fig.~\ref{fig:Bpol} the
%
``poloidal'' component $\bm{B}_p=(B_r,B_z)$ of the magnetic field. The
toroidal component of the magnetic field is maximal at $(20.8/m_V,0)$ and
vanishes on the boundary of the spindle torus, where the field is completely
poloidal. The poloidal field is maximal at the origin
and zero at $(27.4/m_V,0)$. As it was already shown in Fig.~\ref{fig:Btang},
$|\bm{B}_p|=B_c$ along the boundary of the spindle torus. In short terms, the poloidal
component of the magnetic field wraps around the toroidal component, giving rise to the non-zero
helicity of the magnetic field (which is the same as Chern-Simons number for Abelian fields).

\begin{figure}[tbh]
\includegraphics{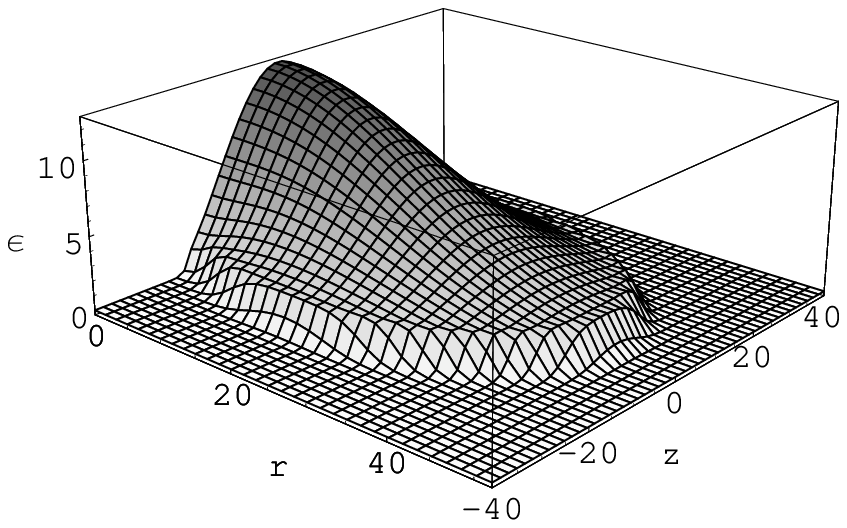}
\caption{\label{fig:Edenscont} 3D plot of the energy density for the soliton with $N_{CS}=12'000$
and $\beta=1$ (in units of the vacuum energy density $\lambda v^4$). The maximum of the energy
density is at the origin and is equal to
$\mathcal{E}_{max}\approx 12.56\ \lambda v^4=6.28\ B_c^2$.}
\end{figure}

Finally we show in Fig.~\ref{fig:Edenscont}
%
the energy density of the soliton ($N_{CS}=12'000$ and $\beta=1$). The different parts of the energy are
\begin{eqnarray}
E_{\text{magn}}&=&\int\frac{\bm{B}^2}{2}\ d^3x\approx 1.00158\cdot 10^5\ m_V/g^2\ ,\\
\label{eq:Ekin}
E_{\text{D}}&=&\int\left[(\bm{\nabla}\phi)^2+g^2\bm{A}^2\phi^2\right]\ d^3x\approx 3.985\cdot 10^3\ m_V/g^2\ ,\\
E_{\text{V}}&=&\int\lambda(\phi^2-v^2)^2\ d^3x\approx 3.2123\cdot 10^4\ m_V/g^2\ .
\end{eqnarray}
The virial theorem \eqref{eq:Derricks relation} is very well satisfied
($E_{\text{magn}}=3E_{\text{V}}+E_{\text{D}}$ to an accuracy of less
than $1\%$). The part $E_{\text{D}}$ of the energy is a surface
term and is negligible for sufficiently large Chern-Simons number.

\subsection{Solitons for different scalar potentials}
In section~\ref{sec:symmPot energy}, we have seen that for a scalar potential
like the one shown in Fig.~\ref{fig:symmPot}, the energy of anomalous solitons
is $E\sim N_{CS}^{2/3}$.

\begin{figure}[tbh]
\includegraphics{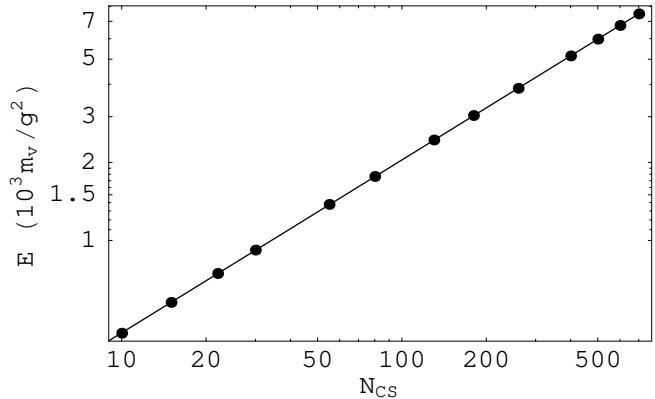}
\caption{\label{fig:EtestNCS} Logarithmic plot of the energy as a function of $N_{CS}$
for anomalous solitons obtained with the potential \eqref{eq:symm-potential}. The dots show
the values of our numerical solutions and the solid line is the best fit of the
form $E=aN_{CS}^{2/3}$, which gives $a\approx 94.866\ m_V/g^2$.}
\end{figure}

\begin{figure}[tbh]
\includegraphics{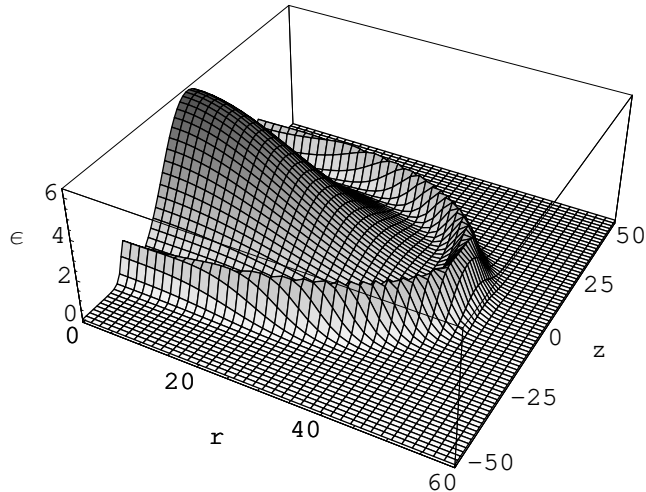}
\caption{\label{fig:Edenstest} The energy density of the soliton with $N_{CS}=500$ and $\widetilde{\beta}=1$ in units
of $\gamma v^4/16$ (which is the potential barrier between the two minima of the potential).}
\end{figure}

In order to verify this conjecture, we also solved the field
equations for the potential
\begin{equation}
\label{eq:symm-potential}
V(\phi)=\gamma(\phi-v)^2\phi^2\ .
\end{equation}
If we again use dimensionless fields and coordinates as before, the field
equations \eqref{eq:hat} for the potential \eqref{eq:symm-potential} read
\begin{subequations}
\label{eq:field eqs symmPot}
\begin{align}
\bm{\nabla}^2\phi-\frac{1}{2}\bm{A}^2\phi
-\frac{\widetilde{\beta}}{4}\left[\phi\left(\phi-1\right)^2+\phi^2\left(\phi-1\right)\right]&=0\ ,\\
\bm{\nabla}\times\left(\bm{\nabla}\times\bm{A}\right)
-2\mu\bm{\nabla}\times\bm{A}+\phi^2\bm{A}&=0\ ,
\end{align}
\end{subequations}
with $\widetilde{\beta}=2\gamma/g^2$. The constraint equation \eqref{eq:dimless NCS}
remains unchanged. We solved Eqs.~\eqref{eq:field eqs symmPot} for $10<N_{CS}<700$
and $\widetilde{\beta}=1$. In Fig.~\ref{fig:EtestNCS}
%
we show the energy of the obtained solutions as a function
of $N_{CS}$. We find
\begin{equation}
E\approx 94.866\ m_V/g^2\ N_{CS}^{2/3}\ ,
\end{equation}
which confirms the claim of section \ref{sec:symmPot energy}. Fig.~\ref{fig:Edenstest}
%
shows the energy density the soliton with $N_{CS}=500$ and $\tilde{\beta}=1$. The main contribution
still comes from the magnetic field, but one can clearly see that there is a contribution
located on the surface, which is the energy of the scalar field. As well in this case, the solitons
are not spherical.

\section{\label{sec:domain shape}Thin Wall Approximation}

\begin{figure}[tbh]
\includegraphics{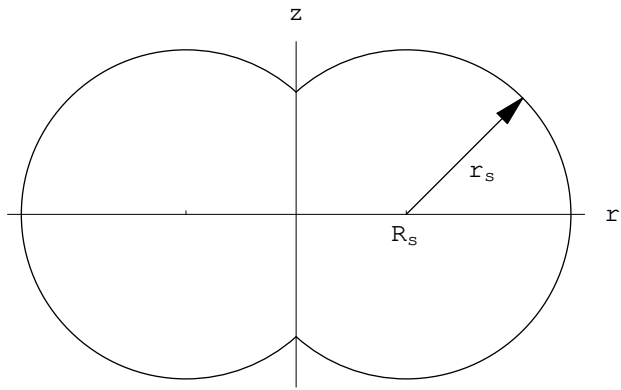}
\caption{\label{fig:spindle} Cross-section of a spindle torus in the $rz$-plane with radii $R_s$ and $r_s$. The spindle
torus is obtained by rotating the circle centered at $(R_s,0)$ with radius $r_s>R_s$ around the $z$ axis.}
\end{figure}

\begin{figure}[tbh]
\includegraphics{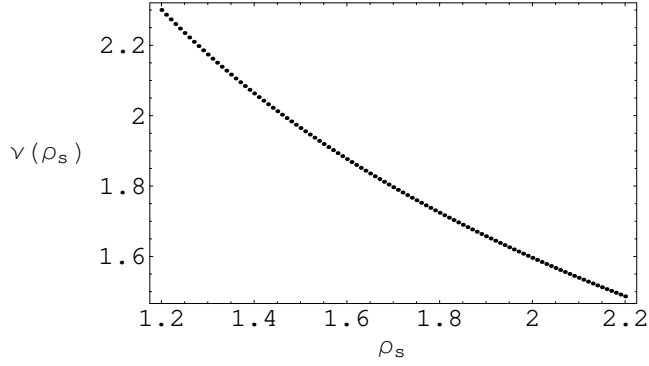}
\caption{\label{fig:eigen} The eigenvalue $\nu$  as a function of the parameter $\vrh_s$.}
\end{figure}

\begin{figure}[tbh]
\includegraphics{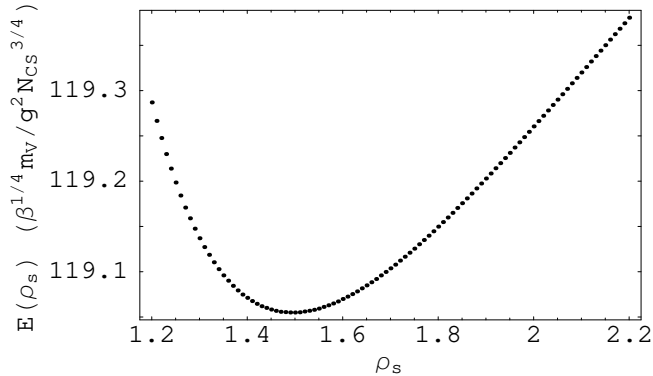}
\caption{\label{fig:Etot} The total energy $E$ as a function of the parameter $\vrh_s$. The minimal energy is at 
$\vrh_s\approx 1.495$ and is $E_{\text{min}}\approx 119.06\ \beta^{1/4}(m_V/g^2)N_{CS}^{3/4}$.}
\end{figure}

\begin{figure}[tbh]
\includegraphics{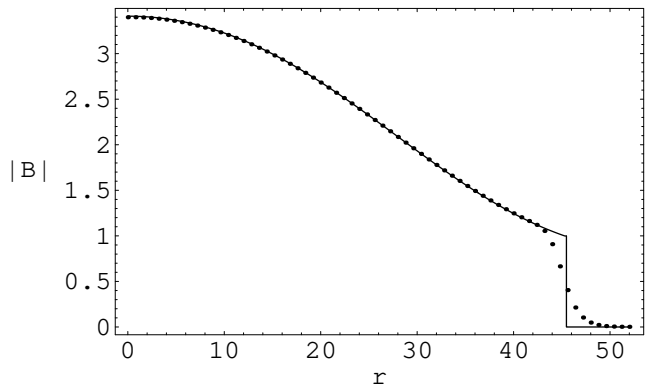}
\caption{\label{fig:BAmpCompare} The amplitude of the magnetic field $|\bm{B}|$ in the plane $z=0$ of
the soliton with $N_{CS}=12'000$ and $\beta=1$. The dotted line shows the solution of the previous
section, the solid line is the solution in the thin wall (macroscopic)
approximation discussed in the present section. The units are $B_c=1$.}
\end{figure}

In this section we construct the anomalous solitons in the thin wall approximation valid for large $N_{CS}$.
In this limit the solitons are described by a compact domain $V$, inside which $\phi=0$. Outside the domain
we have $\phi=v$ and $\bm{A}=0$. In order to determine $\bm{A}$ inside $V$
we have to solve Eq.~\eqref{eq:force-free A} for $\phi=0$,
\begin{equation}
\label{eq:force-free A repeated}
\bm{\nabla}\times\left(\bm{\nabla}\times\bm{A}\right)=2\mu\left(\bm{\nabla}\times\bm{A}\right)\ ,
\end{equation}
where $\bm{A}=0$ on the boundary $\partial V$. In principle, the surface $\partial V$ has also to be determined
by minimization. But because we have found numerically in the previous section,
that the solitons have the shape of a spindle torus for large $N_{CS}$, we take $\partial V$ to be a
spindle torus of fixed size described by two parameters $R_s$ and $r_s>R_s$ (see Fig.~\ref{fig:spindle}).
%
We solve Eq.~\eqref{eq:force-free A repeated} inside the spindle torus as a function of $R_s$ and $r_s$. Then we
minimize the soliton energy with respect to these two parameters. We use dimensionless cylindrical coordinates $(\vrh,\vph,\zeta)$
defined by
\begin{eqnarray*}
x&=&R_s\vrh\cos\vph\ ,\\
y&=&R_s\vrh\sin\vph\ ,\\
z&=&R_s\zeta\ .
\end{eqnarray*}
In these coordinates \eqref{eq:force-free A repeated} simplifies to a system of linear partial differential
equations for the potentials $A_\vrh$, $A_\vph$ and
$A_\zeta$ in the $\vrh\zeta$-plane
\begin{subequations}
\label{eq:rescaled spindle}
\begin{eqnarray}
\label{eq:rescaled spindle a}
\partial_{\zeta}\left(\partial_{\zeta}A_{\vrh}-\partial_{\vrh}A_{\zeta}\right)&=&\nu\ \partial_{\zeta}A_{\vph}\ ,\\
\label{eq:rescaled spindle b}
-\partial_{\zeta}^2 A_{\vph}-\partial_{\vrh}\left[\frac{1}{\vrh}\partial_{\vrh}\left(\vrh A_{\vph}\right)\right]
&=&\nu\ \left(\partial_{\zeta}A_{\vrh}-\partial_{\vrh}A_{\zeta}\right)\ ,\\
\label{eq:rescaled spindle c}
\frac{1}{\vrh}\partial_{\vrh}\left[\vrh\left(\partial_{\zeta}A_{\vrh}-\partial_{\vrh}A_{\zeta}\right)\right]
&=&\nu\ \frac{1}{\vrh}\partial_{\vrh}\left(\vrh A_{\vph}\right)\ ,
\end{eqnarray}
\end{subequations}
with $\nu=2\mu R_s$.\\
Integration of \eqref{eq:rescaled spindle a} and \eqref{eq:rescaled spindle c} gives
\begin{equation}
\nu A_{\vph}=\partial_{\zeta}A_{\vrh}-\partial_{\vrh}A_{\zeta}=R_s B_{\vph}
\label{eq:prop of Aph}
\end{equation}
and using \eqref{eq:prop of Aph} in \eqref{eq:rescaled spindle b} leads to
\begin{equation}
\label{eq:master eq}
\partial_{\vrh}\left[\frac{1}{\vrh}\partial_{\vrh}\left(\vrh A_{\vph}\right)\right]+\partial_{\zeta}^2 A_{\vph}
+\nu^2 A_{\vph}=0\ .
\end{equation}
The Chern-Simons number becomes
\begin{eqnarray}
\label{eq:master constraint}
N_{CS}&=&\frac{g^2}{16\pi^2}\int\bm{A}\cdot\bm{B}\ d^3x\nonumber\\
&=&\frac{g^2}{8\pi^2}\int A_\vph B_\vph\ d^3x\nonumber\\
&=&\frac{g^2\nu}{8\pi^2 R_s}\int A_\vph^2\ d^3x\ ,
\end{eqnarray}
where the second line is found by partial integration and Eq.~\eqref{eq:prop of Aph}
has been used on the third line. We solved Eq.~\eqref{eq:master eq} numerically
with $A_\vph=0$ on the boundary of the spindle torus and under the
constraint \eqref{eq:master constraint}. Because Eq.~\eqref{eq:master eq} is a
linear eigenvalue problem, the solution which minimizes
the energy is the eigenfunction to the lowest positive eigenvalue $\nu^2$, normalized
with Eq.~\eqref{eq:master constraint}. It turns out, that the eigenvalue $\nu$ is a function
of the ratio $\vrh_s=r_s/R_s$, which is shown in Fig.~\ref{fig:eigen}. The total
%
energy of the solution is given by Eq.~\eqref{eq:E(R)} for $R=R_s$. After minimization
with respect to $R_s$, the size of the spindle torus $R_s$ and the energy are given by
(cf. Eqs.~\eqref{eq:domain size} and \eqref{eq:total energy})
\begin{eqnarray}
\label{eq:spindle torus size}
R_s(\vrh_s)&=&\left[\frac{64\pi^2\nu(\vrh_s)}{3I_V(\vrh_s)}\right]^{1/4}\frac{1}{m_V}\left(\frac{N_{CS}}{\beta}\right)^{1/4}\ ,\\
\label{eq:total energy thin wall}
E(\vrh_s)&=&\sqrt{2}\left[\frac{16\pi^2\nu(\vrh_s)}{3}\right]^{3/4}\left[I_V(\vrh_s)\beta\right]^{1/4}\frac{m_V}{g^2}N_{CS}^{3/4}\ ,
\end{eqnarray}
where the shape factor for the spindle torus is
\begin{equation}
I_V(\vrh_s)=\pi^2\vrh_s^2+2\pi\sqrt{\vrh_s^2-1}+\frac{4\pi}{3}\left(\vrh_s^2-1\right)^{3/2}+2\pi\vrh_s^2\arcsin\left(\frac{1}{\vrh_s}\right)\ .
\end{equation}
In Fig.~\ref{fig:Etot} we show the function $E(\vrh_s)$,
%
which can be minimized numerically with respect to $\vrh_s$. There is a minimum at $\vrh_{s}\approx 1.495$, for which we get
\begin{eqnarray}
\label{eq:minimal spindle size}
R_s&\approx&1.742\ \frac{1}{m_V}\left(\frac{N_{CS}}{\beta}\right)^{1/4}\ ,\\
\label{eq:minimal energy}
E&\approx&119.065\ \beta^{1/4} \frac{m_V}{g^2} N_{CS}^{3/4}\ .
\end{eqnarray}
These results are in excellent agreement with the results of the
previous section, where we obtained
\begin{eqnarray}
R_s&\approx&1.726\ \frac{1}{m_V}\left(\frac{N_{CS}}{\beta}\right)^{1/4}\ ,\\
E&\approx&118.826\ \beta^{1/4} \frac{m_V}{g^2} N_{CS}^{3/4}\ .
\end{eqnarray}
The macroscopic solutions discussed in the present section coincide with the solutions of the previous
section for large $N_{CS}$. To illustrate this, we compare in Fig.~\ref{fig:BAmpCompare} the
%
magnetic fields of the macroscopic solution and the solution of the previous section.

\section{\label{sec:summary}Conclusions}
In this article we studied anomalous Abelian solitons, a class of solitons of the chiral Abelian Higgs model
found by Rubakov and Tavkhelidze. These solitons carry
a non-zero Chern-Simons number $N_{CS}$ and they are stable for large $N_{CS}$.
We showed that their energy-versus-fermion-number ratio is given by $E\sim N_{CS}^{3/4}$ or
$E\sim N_{CS}^{2/3}$ depending on the structure of the scalar potential.
For the former case we derived a lower bound for the soliton energy,
$E\geq c N_{CS}^{3/4}$, where $c$ is a constant depending on the coupling constants and the VEV of the
Higgs field.

We discussed the structure of anomalous solitons by solving numerically the field equations
for both the gauge and the Higgs field. It turned out that the solitons are not spherically symmetric.
For large Chern-Simons number they have the shape of a spindle torus. In this limit we also
constructed the solutions in the thin wall (macroscopic) approximation.

At the moment it is not clear, if (metastable) anomalous solitons exist in the standard model.
It will be interesting to extend the present study to the full electroweak theory with
$SU(2)\times U(1)$ gauge group.

\begin{acknowledgments}
This work was supported by the Swiss National Science Foundation. We thank Andrei Gruzinov for helpful
discussions.
\end{acknowledgments}

\appendix
\section{\label{App:A}Inequalities}
In this appendix we review the inequalities from functional analysis, which have been used in Sec.~\ref{sec:lower bound} to
derive the lower bound on the energy of the anomalous solitons.
\subsection{\label{sec:Höldersche Ungleichung}H\"older inequality}
Let $\Omega$ be an open subset of $\mathbb{R}^d$ and $p,\ q\geq 1$ such that
\begin{equation*}
\frac{1}{p}+\frac{1}{q}=1\ .
\end{equation*}
Then, for all functions $f\in L^p(\Omega)$ and $g\in L^q(\Omega)$, the product $fg\in L^1(\Omega)$ satisfies
the H\"older inequality,
\begin{equation}
\label{eq:Höldersche Ungleichung}
\lVert fg\rVert_{L^1}\leq\lVert f\rVert_{L^p}\lVert g\rVert_{L^q}\ ,
\end{equation}
where the norm on $L^p(\Omega)$ is defined by
\begin{equation*}
\lVert f\rVert_{L^p}:=\left(\int\limits_{\Omega} \lvert f\rvert^p\ dx\right)^{1/p}\ .
\end{equation*}
Note that for $p=q=2$ the H\"older inequality reduces to the Cauchy-Schwarz inequality. For
vector fields $\bm{u}(\bm{x}):\Omega\rightarrow\mathbb{R}^{d}$ the $L^p$-norm
is defined by
\begin{equation*}
\lVert\bm{u}\rVert_{L^p}=\left(\int\limits_{\Omega}(\bm{u},\bm{u})^{p/2}\ dx\right)^{1/p}\ ,
\end{equation*}
where $(\cdot,\cdot)$ denotes the scalar product of vectors in $\mathbb{R}^d$. Thus for vector
fields the H\"older inequality reads
\begin{equation}\label{eq:Holder inequality for vectors}
\Biggl\vert\int\limits_{\Omega}(\bm{u},\bm{v})\ dx\Biggr\vert\leq\lVert\bm{u}\rVert_{L^p}\lVert\bm{v}\rVert_{L^q}\ .
\end{equation}
\subsection{\label{A.2}``Modified'' H\"older inequality}
As a consequence of \eqref{eq:Höldersche Ungleichung} we have for $s>1$ that
\begin{equation}
\label{eq:modified Hölder}
\lVert f\rVert_{L^s}^s\leq \lVert f\rVert_{L^1}^{2-s}\lVert f\rVert_{L^2}^{2s-2}\ ,\quad\forall f\in L^s(\Omega)\ .
\end{equation}
The proof of \eqref{eq:modified Hölder} is very simple: defining the functions $g$ and
$h$ by
\begin{align*}
g&=|f|^{2-s}\ ,\\
h&=|f|^{2s-2}\ ,
\end{align*}
we get
\begin{align*}
\lVert f\rVert_{L^s}^s&=\lVert gh\rVert_{L^1}\\
&\leq \lVert g\rVert_{L^{1/(2-s)}}\lVert h\rVert_{L^{1/(s-1)}}\\
&=\left(\int\limits_\Omega |g|^{1/(2-s)}\ dx\right)^{2-s}\left(\int\limits_\Omega |h|^{1/(s-1)}\ dx\right)^{s-1}\\
&=\left(\int\limits_\Omega |f|\ dx\right)^{2-s}\left(\int\limits_\Omega|f|^2\ dx\right)^{\frac{2s-2}{2}}\\
&=\lVert f\rVert_{L^1}^{2-s}\lVert f\rVert_{L^2}^{2s-2}\ ,
\end{align*}
where \eqref{eq:Höldersche Ungleichung} has been used on the second line.
\subsection{\label{sec:Sobolev Ungleichung}Gagliardo-Nirenberg-Sobolev inequality}
The Sobolev spaces $W^{n,p}(\Omega)$, where $p\geq 1$ and $n\in\mathbb{N}$, are the
normed spaces of functions defined by
\begin{equation*}
W^{n,p}(\Omega)=\Bigl\{f\in L^p(\Omega)\big\vert\ \forall\alpha\in\mathbb{N}^d,\ |\alpha|\leq n:\partial^\alpha_x f\in L^p(\Omega)\Bigr\}\ .
\end{equation*}
For $1\leq p<d$ we define the Sobolev conjugate of $p$ by
\begin{equation*}
p^*=\frac{dp}{d-p}\ .
\end{equation*}
Then, there is a constant $c(d,p)$, such that
\begin{equation}
\label{eq:Sobolev Ungleichung}
\lVert f\rVert_{L^{p^*}}\leq c(d,p)\ \lVert Df\rVert_{L^p}\ ,\quad\forall f\in W^{1,p}(\Omega)\ ,
\end{equation}
where $D$ is the weak derivative.~\eqref{eq:Sobolev Ungleichung}
is known as Gagliardo-Nirenberg-Sobolev (GNS) inequality. For
$\Omega=\mathbb{R}^d$ the best possible constant in \eqref{eq:Sobolev Ungleichung} is
\begin{equation*}
c(d,p)=
\begin{cases}
\frac{1}{d}\left(\frac{d}{|S^{d-1}|}\right)^{1/d}&p=1\\
\frac{p-1}{d-p}\left[\frac{d-p}{d(p-1)}\right]^{1/p}
\left[\frac{\Gamma(d+1)}{\Gamma\left(d/p\right)\Gamma\left(d+1-d/p\right)|S^{d-1}|}\right]^{1/d}&p>1\ .
\end{cases}
\end{equation*}
Here, $|S^{d-1}|$ denotes the surface area of the $(d-1)$-dimensional unit sphere,
\begin{equation*}
|S^d|=2^d\pi^{d/2}\frac{\Gamma(d/2)}{\Gamma(d)}\ .
\end{equation*}
In Sec.~\ref{sec:lower bound} we have used \eqref{eq:Sobolev Ungleichung} for continuosly differentiable functions
with compact support and $(d,p)=(3,2)$.
In this case the weak derivative is just the ordinary gradient and the GNS inequality reads
\begin{equation}
\label{eq:Sobolev inequality}
\lVert f\rVert_{L^6}\leq c(3,2)\ \lVert \bm{\nabla}f\rVert_{L^2}
=\frac{1}{\sqrt{3}}\left(\frac{2}{\pi}\right)^{2/3}\ \lVert \bm{\nabla}f\rVert_{L^2}\ ,\quad\forall f\in C^1_c(\mathbb{R}^3)\ .
\end{equation}
\section{\label{App B} Numerical procedure}
In this appendix we explain the numerical procedure which was used to obtain numerical solutions of the PDEs \eqref{eq:sys incl phi}
subject to the constraint \eqref{eq:sys constr} and boundary conditions \eqref{eq:sys boundary conditions}.
Solutions are calculated on the rectangular box $[0,L]\times[-Z,Z]$ in the $rz$-plane. The dimensions of the box
$L$ and $Z$ have to be at least a few times $N_{CS}^{1/4}$, since the expected size of the soliton is $R\sim N_{CS}^{1/4}$.
The box is divided into a grid of $N_r\times N_z$ points $(r_i,z_i)$ given by
\begin{align*}
r_i&=h_r(i-1)\ ,\qquad\qquad\text{for}\quad i=1,\ldots,N_r\ ,\\
z_j&=\frac{h_z}{2}(2j-1-N_z)\ ,\quad\text{for}\quad j=1,\ldots,N_z\ ,
\end{align*}
where $h_r=L/(N_r-1)$ and $h_z=2Z/(N_z-1)$. Because the coordinates $(r,z)$ are in units of $1/m_V$ and we expect the fields
to vary on scales of order $1/m_H$ and $1/m_V$, $h_r$ and $h_z$ have to be smaller than $\beta^{1/2}$ to ensure a sufficient
resolution of the solutions. Therefore, the number of grid points has to be increased
going either to larger $N_{CS}$ or to lower $\beta$. As an example for $N_{CS}=12'000$
and $\beta=1$, we were using the rectangular box $[0,52]\times[-40,40]$ with a resolution of $0.8/m_V$, which lead to a grid
with $6'666$ points.

The derivatives in the PDEs \eqref{eq:sys incl phi} are approximated by finite differences in a straightforward way,
\begin{align*}
f(r_i,z_j)&\rightarrow f_{i,j}\ ,\\
(\partial_r f)(r_i,z_j)&\rightarrow\frac{1}{2h_r}\left(f_{i+1,j}-f_{i-1,j}\right)\ ,\\
(\partial_z f)(r_i,z_j)&\rightarrow\frac{1}{2h_z}\left(f_{i,j+1}-f_{i,j-1}\right)\ ,\\
(\partial_r^2 f)(r_i,z_j)&\rightarrow\frac{1}{h_r^2}\left(f_{i+1,j}-2f_{i,j}+f_{i-1,j}\right)\ ,\\
(\partial_r\partial_z f)(r_i,z_j)&\rightarrow\frac{1}{4h_rh_z}\left(f_{i+1,j+1}-f_{i+1,j-1}-f_{i-1,j+1}+f_{i-1,j-1}\right)\ ,\\
(\partial_z^2 f)(r_i,z_j)&\rightarrow\frac{1}{h_z^2}\left(f_{i,j+1}-2f_{i,j}+f_{i,j-1}\right)\ ,
\end{align*}
where $f(r_i,z_j)$ stands for one of the fields $A_r$, $A_\vph$, $A_z$ or $\phi$ at the position $(i,j)$ on the grid. Using these
replacement rules the PDEs reduce to a system of non-linear equations on the internal points of the grid (i.e.
$i=2,\ldots, N_r-1$ and $j=2,\ldots,N_z-1$). On the axis of symmetry $r=0$ the finite difference approximation
for the PDEs requires special care. The regularity conditions \eqref{eq:reg cond} at $r=0$ for the fields $A_r$ and $A_\vph$ imply
\begin{align}
A_{r,1,j}&=0\ ,\\
A_{\vph,1,j}&=0\ ,
\end{align}
for $j=2,\ldots,N_z-1$. For discretizing Eqs.~\eqref{eq:sys ph} and \eqref{eq:sys az} at $r=0$ we have to use the replacements 
\begin{align*}
\frac{1}{r}\partial_r\left(r\partial_r\phi\right)_{r=0}=2\partial_r^2\phi\Big|_{r=0}&\rightarrow \frac{4}{h_r^2}\left(\phi_{2,j}-\phi_{1,j}\right)\ ,\\
\frac{1}{r}\partial_r\left(r\partial_r A_z\right)_{r=0}=2\partial_r^2 A_z\Big|_{r=0}&\rightarrow \frac{4}{h_r^2}\left(A_{z,2,j}-A_{z,1,j}\right)\ ,\\
\frac{1}{r}\partial_r\left(r\partial_z A_r\right)_{r=0}=2\partial_r\partial_z A_r\Big|_{r=0}
&\rightarrow\frac{1}{h_rh_z}\left(A_{r,2,j+1}-A_{r,2,j-1}\right)\ ,\\
\frac{1}{r}\partial_r\left(rA_\vph\right)_{r=0}=2\partial_r A_{\vph}\Big|_{r=0}&\rightarrow \frac{2}{h_r}A_{\vph,2,j}\ .
\end{align*}
The system of non-linear equations is completed by the boundary conditions at $r=L$ and $z=\pm Z$, which read
\begin{subequations}
\label{eq:bc}
\begin{align}
\phi_{i,1}&=\phi_{i,N_z}=1\quad\text{and}\quad\bm{A}_{i,1}=\bm{A}_{i,N_z}=0\quad\text{for}\quad i=1,\ldots,N_r\ ,\\
\phi_{N_r,j}&=1\qquad\qquad\text{and}\qquad\bm{A}_{N_r,j}=0\qquad\text{for}\quad j=1,\ldots,N_z\ .
\end{align}
\end{subequations}
Finally, the constraint equation \eqref{eq:sys constr} is given by
\begin{equation}
\label{eq:ConstrEqn}
N_{CS}=\frac{h_rh_z}{8\pi}\sum_{i=2}^{N_r-1}\sum_{j=2}^{N_z-1}h_r(i-1)A_{\vph,i,j}
\left(\frac{A_{r,i,j+1}-A_{r,i,j-1}}{2h_z}-\frac{A_{z,i+1,j}-A_{z,i-1,j}}{2h_r}\right)\ .
\end{equation}
Including the constraint equation \eqref{eq:ConstrEqn} we obtain a system of $4N_rN_z+1$ non-linear equations for the 
unkowns $\phi_{i,j}$, $A_{r,i,j}$, $A_{\vph,i,j}$, $A_{z,i,j}$ and the Lagrangian multiplier $\mu$. This system is solved
using a standard Newtonian algorithm for non-linear equations. In every step of the Newton iteration, one has to solve
a system of $4N_rN_z+1$ linear equations.

\bibliography{AnAbSoliton}
\bibliographystyle{h-elsevier}
\end{document}